# Size Does Not Matter – In the Virtual World. Comparing Online Social Networking Behavior with Business Success of Entrepreneurs

Gloor, P. A., Woerner, S. L., Schoder, D., Fischbach, K., & Fronzetti Colladon, A.

# Size Does Not Matter – In the Virtual World.

# Comparing Online Social Networking Behavior with Business Success of Entrepreneurs

Peter A. Gloor [a], Stephanie Woerner [b], Detlef Schoder [c], Kai Fischbach [d], Andrea Fronzetti Colladon [e1]

**Abstract**

We explore what benefits network position in online business social networks like LinkedIn might confer to an aspiring entrepreneur. We compare two network attributes, size and embeddedness, and two actor attributes, location and diversity, between virtual and real-world networks. The promise of social networks like LinkedIn is that network friends enable easier access to critical resources such as legal and financial services, customers, and business partners. Our setting consists of one million public member profiles of the German business networking site XING (a German version of LinkedIn) from which we extracted the network structure of 15,000 startup entrepreneurs from twelve large German universities. We find no positive effect of virtual network size and embeddedness, and small positive effects of location and diversity.

# 1. Introduction

Networks and relationships are central to the entrepreneurial process (Dubini and Aldrich, 1991); they can be antecedents of firms' success (MacMillan, 1983), affect early performance and help in accessing strategic knowledge to recognize entrepreneurial opportunities (Baum et al., 2000; Shane and Venkataraman, 2000). Innovative companies connect more with each other through online networking, operate more efficiently and leverage social contacts to improve business performance (Brüderl and Preisendörfer, 1998; Ellison et al., 2007). Usually, CEOs of small and medium enterprises perceive online social networks as beneficial to their companies (Zaglia et al., 2015). The benefits of social relationships have been investigated both at the individual (Uzzi and Spiro, 2005) and firm level (Uzzi, 1996; 1997). In more recent research, the relationship between social network sites, such as LinkedIn, and entrepreneurial success has been explored (Song and Vinig, 2012). We see examples, like executive recruiter Ron Bates (who has 43,000 LinkedIn friends according to his LinkedIn profile[1]), of business people compiling large networks of business contacts. Previous studies reported that business entrepreneurs use social networking sites to develop, build and maintain relationships with potential funders, clients or mentors and to find candidates for job openings (Song, 2015). Social networking sites are also useful for information seeking, to cultivate relationships with people met in person, to participate in business communities and to engage in business development (Baron and Markman, 2003; Ferro, 2015). In this paper, we explore what the value might be of these large networks of online ties, focusing on the German community of Xing users. German knowledge workers use social networking sites to stay in touch with people they know, to manage their contacts, to search for experts and learn

---

[1] Ron Bates is a Managing Principal and Search Consultant with Executive Advantage Group and has according to his own claim made in a CNN interview among the largest number of friends on LinkedIn. See http://www.linkedin.com/in/ronbatesprofile and http://www.edition.cnn.com/video/#/video/world/2011/05/18/qmb.linked.ipo.price.cnn (retrieved March 10, 2013)

about specific topics (Ferro and Zachry, 2014; Richter and Koch, 2008). We investigate the value of online ties for entrepreneurs. While the value of real-world networking is well understood, we do not really know if having many friends in LinkedIn, Facebook and other social networking sites is good for business success. While LinkedIn is undoubtedly useful for certain types of users, such as executive recruiters like Ron Bates, it is not obvious if having thousands of LinkedIn friends really helps the startup entrepreneur.

**How virtual ties differ from real world networks?**

A large body of research has shown that Internet use fosters social networks size and diversity (Chen, 2013; Hampton, 2011; Hampton et al., 2011; Zhao, 2006). There are conceptual differences between offline and online ties. While real-world ties are well understood (Lee, 2008; Moran, 2005; Porter et al., 2005; Raz and Gloor, 2007; Schilling and Phelps, 2005; Sparrowe et al., 2001; Uzzi, 1996) and researchers have thoroughly explored the respective advantages of strong and weak ties for business success, much less research has focused on the properties of virtual networks (Aral and Walker, 2012; Centola, 2010; Wu, 2003). In research comparing offline and online ties, Kuhn et al. (2016) explore the process of advice seeking of entrepreneurs and stress the potential of online communication and of weak ties, finding an association of online communication with business growth. In their study, they also find that women and younger business owners are more inclined to rely on social media and online forums to access advice from other entrepreneurs. However, they do not make a clear distinction among online sources of advice, while there might be a difference in support coming from specialized forums and more general online sources, like Facebook or LinkedIn. While the strength of real-world ties can be measured through the associated social capital, this is much harder for virtual ties, as a friendship link in LinkedIn or Facebook is generally only a marker, with little information about the strength of the tie (De Meo et al., 2014).  In addition, there are different types of virtual ties. Often it is not just a matter of communication – which might come from offline encounters or from friendship links on social media – but what really matters is the quality of the knowledge or advice which are shared and the associated communication behavior (Allen et al., 2016).

For the purpose of our study we distinguish between three types of ties: Some ties exist only online, some are only offline, and some might be both.

*Only real-world ties:* The usefulness of having the right types of ties – strong or weak – has become increasingly apparent in the real world (Battilana and Casciaro, 2012). Real world links are stronger than purely virtual ones, in that the two people are likely to have met face-to-face at least once and share some mutual interest or experience. In addition, these ties, independent of strength, are by definition, remembered by the individual.

*Only online ties:* Today there are many LinkedIn or Facebook users with more than 1000 friends. According to Wolfram (2013) as of April 2013 on Facebook the average user has 342 friends, with millions of Facebook users over the 1000 friends mark. It is likely that many of these "friends" are people they have met only once or are friends of friends, people without a face to a name. Many Facebook or LinkedIn users send out friend requests to people they have only met in passing and accept any friendship request that they receive; assembling a large number of friends on Facebook or LinkedIn is sometimes seen as a badge of honor. While social ties in the real world are invisible and have to be extracted through surveys and interviews (Hoang and Antoncic, 2003), virtual ties can be created easily, by request, accessed with little effort, and are visible to all. It is comparatively easy to amass huge numbers of online ties, raising the question of whether having many virtual ties conveys any competitive advantage. Preferred access to critical resources gives an actor in a business setting a direct competitive advantage (Pfeffer and Salancik, 1978); yet intuition suggests that it is hard to benefit from a prominent network position when large numbers of people have access to the same position and resources. Nevertheless, it could still be that prolific link collectors derive some value from their virtual tie by, for instance, sharing status updates and requests on their LinkedIn wall.

*Online tie as a marker for a real-world tie:* Frequently, link-collectors — people who initiate many requests for "friends" — do not personally know the person with whom they are connected. Online ties, are, however, sometimes a marker of a relationship in the real world. On Facebook, these might be links between family members. In LinkedIn, a virtual tie might be a marker of a real-world relationship, if the two individuals sharing the link have worked for the same company, or are alumni of the same university. However, if an individual has more than a few hundred ties in an online network, we conjecture a large number of them are exclusively virtual without a corresponding real-world tie. LinkedIn ties are different from ties extracted from e-mail networks. While the number of e-mails exchanged between two actors is a proxy for the weight of the link, LinkedIn links are dichotomous, they either exist or they do not, with little indication about the strength of the tie.

In this paper, we focus on virtual ties of the weakest form, collected from XING, a German online business social networking site such as LinkedIn. Our aim is to investigate if entrepreneurs will be able to derive any competitive advantage from their online virtual business networks through network size, network position, and friend attributes such as location or profession (Figure 1).

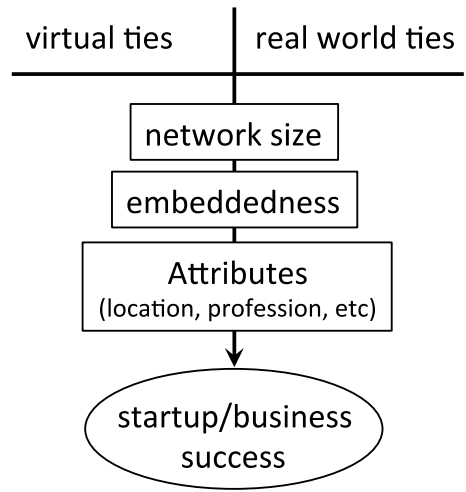

Figure 1. Research Framework

## 2. Do Successful Entrepreneurs Have Distinctive Online Networks

Researchers have done studies on the relevance and importance of networks in entrepreneurship for more than 20 years (e.g., Hoang and Antoncic, 2003; Strobl, 2014; Weber and Kratzer, 2013;). It has been shown that networks are central to the entrepreneurial process (Dubini and Aldrich, 1991). Entrepreneurs with many social ties have been shown to have clear advantages (Schilling and Phelps, 2005; Watson, 2007). Social ties and networks are especially important in high-technology ventures (Elfring and Hulsink, 2003; Stam and Elfring, 2008; Vissa and Chacar, 2009). Allen et al. (2016) investigated the impact of communication on innovation capabilities of biotech start-ups, finding that the quality of the exchanged knowledge matters more than physical proximity. What seems to be most important is communication intensity and recollection by others. Recently, online social network communication, such as blogs, Facebook, MySpace, LinkedIn, and Twitter have become a major means of staying in touch with friends and business partners, complementing, and even substituting for, established communication channels such as e-mail and phone (Boyd, 2008; Miller and Tucker, 2013). This increase in online social communications leads us to explore whether the business benefits of social ties extend to online social networks.

As the value of networking is heavily touted in business schools and entrepreneurship programs, we expect entrepreneurs to have more online friends than non-entrepreneurs (Kumar, 2007; Mehra et al., 2006), leading to larger online network size. What is less clear is if there is any value for the entrepreneur in having so many online friends, and if there are differences in network structure and attributes between successful and non-successful entrepreneurs. While most online friends of an entrepreneur may be purely virtual without a corresponding real-word tie, they may still provide access to some specific types of resources. In particular, we speculate that networks differ in three aspects: (1)

number of friends, (2) network position, and (3) network composition, and that we see these distinct differences predicting success among the entrepreneurs.

As online social networking requests are frequently accepted without knowing the requester, it is difficult to measure the strength of a business social networking tie. We decided to use membership in the same alumni network as a proxy for social capital and geographic proximity, as students attending the same university at approximately the same time can be assumed to have been in closer geographic proximity, and have a higher chance of having physically met.

In our study, we analyze a dataset from XING, the German equivalent of LinkedIn. In a related study a LinkedIn network was analyzed for a group of entrepreneurs (Gloor et al., 2013). However, LinkedIn only allows collection of direct friendship network of participating entrepreneurs. In this project we obtained the full network of  German entrepreneurs on XING , which would be impossible to obtain for LinkedIn. XING and LinkedIn share many similarities:  they have the same purpose of business networking, they were both launched in the same year (2003) and they also apply a similar general taxonomy (Pallis et al., 2011). We therefore argue that our results, while collected for Germany, are also valid for the US and LinkedIn.

While in the US alumni from top universities such as Harvard or MIT have a clear advantage, and are members of a strong "old boys" network (Cohen et al., 2008; Lerner et al., 2008;), this is less clear in the German universities. We therefore include in our discussion analogies and differences of the German university system compared to the US system. In Germany, there is a network of mostly free public universities with LMU Munich and University of Cologne leading the list in terms of size, with well over 40,000 students each. These universities are funded by the state, and offer degrees in all fields of science and the arts. At the other extreme, and most relevant for entrepreneurs, are small private universities like WHU (Otto Beisheim School of Management) with less than 800students and EBS (European Business School) with 1500 students, which exclusively concentrate on business education up to the doctoral level. While the public universities charge only a few hundred Euros per year for tuition, a bachelor's degree at EBS or WHU will cost over 30,000 Euros[2]. In a study by Wirtschaftswoche, it was found that the return to getting an education at a private university can be substantial, with WHU graduates getting the highest starting salaries of students at all German universities[3]. It therefore appears that investing into a business degree at EBS or WHU, similar to a degree from an Ivy League business school, might have the potential to pay off financially later in professional life.

In addition, alumni from "elite" universities – with higher organizational distinctiveness and prestige – tend to show more alumni identity (Mael and Ashford, 1992) compared to alumni from mass universities. It could be, however, that the large, but well-financed public universities such as LMU could offer their alumni broader, bigger networks. Therefore, we include in our sample alumni from large German mass universities as well as alumni from the smaller private universities EBS and WHU.

Online social networking ties in LinkedIn or XING provide little preferential access and embedded social capital (Sparrowe et al., 2001), exhibiting characteristics more typical of low-density, low-identity, low-clustering coefficient weak ties. Weak ties can be beneficial because they bridge to distant locales (Granovetter, 1973) and give access to diverse information. Even if strong ties can be vital in the process of small-business formation, in building trust and for problem solving issues, the lack of weaker bridging ties between different communities may prevent entrepreneurs from getting access to important information and social resources (Song, 2015). Accordingly, weak ties are also essential in the start-up phase of each business, (Song, 2015). Having many friends in LinkedIn might also have a positive signaling effect – similar to the feathers of the peacock, where the male peacock with the most brilliant feathers has an evolutionary advantage as females see him as the most attractive. In the online social networking context more friends might signal higher access to social capital, and a willingness to promote oneself useful for starting a new business. However, there can be some cost involved in creating

---

[2] http://de.wikipedia.org/wiki/WHU_–_Otto_Beisheim_School_of_Management

[3] http://www.spiegel.de/spiegelspecial/a-337421.html (Spiegel online, 18.1.2005, retrieved March 10,2013)

these weak tie networks, scouring the Internet and business networking events for possible online friends and sending out the XING friendship requests. This time might be better spent building the business. We therefore speculate that online social business network size and embeddedness will not be a significant predictor for long-term business success of an entrepreneur.

For entrepreneurs, diverse ties to high-status alters are particularly helpful (Lin et al., 1981). The influence of external advice networks has been shown to be associated with high growth ventures (Vissa and Chacar, 2009). Preferred access to legal or financial advice, as well as to capital, might assist entrepreneurs unable to rely on a high-value private-university alumni network in growing their business (Roure and Keeley, 1990). We therefore hypothesize that having the right friends might help entrepreneurial success and that successful entrepreneurs will make sure to obtain online access to preferred resources such as lawyers, bankers, and venture capitalists (Romanelli, 1989; Roure and Keeley, 1990;). We also assume that this relationship holds the other way around: bankers, lawyers, and venture capitalists will look for preferred access to successful entrepreneurs. Either way, a successful entrepreneur will have more virtual ties to bankers, lawyers, and venture capitalists.

## 3. Data Analysis

The first step involved crawling the publicly accessible profiles of people on XING (http://www.xing.com). XING is the leading German language business networking Web site, similar to LinkedIn in the U.S. People on XING have the option of either hiding or disclosing their profile to the outside world, as well as of hiding or disclosing their friends. Our analysis is restricted to people who chose to make their profile data publicly accessible, including a list of their friends. We believe that this is a representative sample of German startup entrepreneurs for several reasons. First, most adult Germans are active online, 78% in 2013 according to "Forschungsgruppe Wahlen" (2013). Second, as the default in XING is to publicly display the profile and also show the friendship links, the majority of people registering in XING before 2009 chose that option – awareness of this issue in Germany rose significantly only at the end of 2009 after some high-profile incidents[4] well after we had finished our data collection. Third, each link was touched twice by our crawler, so even if one person had hidden the profile or the link, the crawler got a second chance to collect it from the other side. Figure 2 shows the degree distribution of the XING members in our sample.

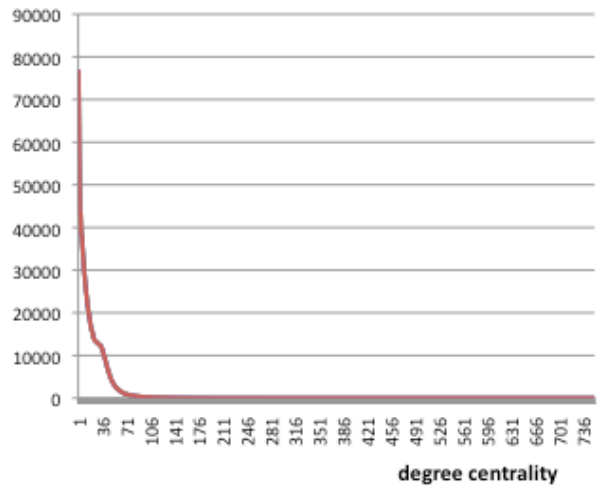

*Figure 2. Degree distribution of XING members in sample, y-axis is number of people, x-axis is their*

*degree centrality, i.e. number of direct XING friends (N=981,385)*

We took this dataset and segmented it into networks using the German university system as the key descriptor. In Germany in 2013, out of the 108 universities, nine have more than 40,000 students, and another between 20,000 and less than 40.000 students. They encompass a comprehensive set of disciplines. Private universities are smaller and typically have student bodies of less than 5,000 students;

many of these smaller institutions focus on one discipline (such as medicine, arts and sciences, or engineering) and do not offer doctoral degrees. We randomly selected networks of ten of the largest public German universities: University of Cologne, HU Berlin, University of Hamburg, University of Hannover, University of Mannheim, LMU Munich, FU Berlin, RWTH Aachen, TU Munich, and University of Karlsruhe. We also included networks of the two major private business schools with programs comparable to those at the public universities: European Business School of Oestrich-Winkel (EBS) and WHU Otto Beisheim School of Management. Including EBS and WHU allowed us to investigate whether there is a difference in startup founder performance between networks based on more exclusive membership (i.e., alumni of private universities), and those more broadly based (i.e., alumni of public German universities). Public universities have to admit anybody fulfilling the admission requirements, while private business schools such as WHU or EBS claim to be highly selective.

---

[4] http://www.spiegel.de/netzwelt/netzpolitik/kopierte-nutzerinfos-datenpanne-draengt-schuelervz-in-die-defensive-a-655820.html

Overall we collected 981,385 user profiles with 10,212,704 links (see Figure 2). According to the XING annual report for 2009, end of 2008 XING had 7 million members and 124 million contacts, i.e. roughly 18 links per member. With our crawler we were able to obtain 10 links per member, i.e. more than half. We assume that the subset of friends we get is a random sample of the full friend list without significant selection bias. In general, we expect entrepreneurs to be more willing to be found on XING and thus allow access to their link data. In addition, our crawler had two chances to collect a link between two people, one chance from each member of the dyad.

In addition to the social network data, we gathered data on the number of registered students and the number of students graduating each year from 2004 to 2007 of the 12 universities (see Table 1).

| University | Total students Ø 2004-07 | Graduates Ø 2004-07 | Alumni (Xing sample) | Founders (Xing sample) | Group Betweenness Centrality | Group Degree Centrality |
|---|---|---|---|---|---|---|
| EBS | 1270 | 285 | 554 | 173 | 0.2015 | 0.0603 |
| FU Berlin | 33646 | 4356 | 6172 | 1608 | 0.1853 | 0.0394 |
| HU Berlin | 29570 | 3683 | 1650 | 383 | 0.0269 | 0.0441 |
| LMU Munich | 43722 | 6025 | 6504 | 2726 | 0.1924 | 0.0342 |
| RWTH Aachen | 29441 | 2960 | 5769 | 1266 | 0.1444 | 0.0259 |
| TU Munich | 21237 | 3740 | 3076 | 1262 | 0.1557 | 0.0189 |
| U Cologne | 45158 | 5019 | 7826 | 2210 | 0.1855 | 0.0427 |
| U Hamburg | 37518 | 4982 | 9128 | 2526 | 0.1418 | 0.0249 |
| U Hannover | 22144 | 2650 | 3500 | 857 | 0.1406 | 0.0220 |
| U Karlsruhe | 17579 | 2089 | 3577 | 821 | 0.1131 | 0.0252 |
| U Mannheim | 11089 | 1380 | 3562 | 826 | 0.1197 | 0.0321 |
| WHU | 444 | 158 | 658 | 196 | 0.2890 | 0.2131 |

*Table 1. Basic data for all 12 universities*

To further focus on the networks of students, entrepreneurs, and executives of startups, we systematically parsed the publicly accessible alumni profiles of these universities. In the profiles we searched for keywords such as "Chief", "Inhaber" (owner), "Besitzer" (owner), "Unternehmer" (entrepreneur), "Jungunternehmer" (junior entrepreneur), "Gesellschafter" (shareholder), "Geschäftsführer" (CEO), "Geschäftsführender" (CEO), "Gründer" (founder), "Teilhaber" (Co-owner), "Enterpriser", "Entrepreneur", and "Startup".

Out of the initial sample, 15,143 profiles were identified (based on the keywords listed above) as founders and entrepreneurs from the 12 universities. These founders had 232,390 links. Another 130,390 profiles were their XING friends, either alumni from the 12 universities or from other organizations.

We randomly selected 80 founders from each university, leading to a total sample of 960 randomly chosen founders and their associated social networks.[5] We created a matched sample by randomly choosing a second sample of 960 non-founders with networks—again 80 from each university—from our full database.

---

[5] We also manually checked the demographics for this sample. 403 of the 960 founders included information about the graduating year in their profile on XING. 71% of them graduated between 1998 and 2008, which puts them in the age group of the online-social-networking-savvy 26 to 36 year olds.

What about entrepreneurs who started a firm but failed? Unsuccessful founders are included among the entrepreneurs in our sample through the longitudinal design of our study. These were the entrepreneurs who were in the roster in the first evaluation round, but fell out in the second round after 18 months.

**Network centrality and betweenness**

We computed the *degree centrality* and *betweenness centrality* for each actor in the 12 alumni networks. We calculated the betweenness of each actor in the full network with 981,385 actors, as well as in a pruned network where we only included the top 100,000 most connected users (having on average more than 43 friends) to remove casual users. We also calculated the number of ties to fellow alumni (the *in-group*) and the ties that are to people who are not fellow alumni (the *out-group*). We chose these two metrics as particularly appropriate for online social networks, representing accessibility (degree) and information flow (betweenness) (Everett and Borgatti, 2005; Wasserman and Faust, 1994). (See Table 2 for all variables and the associated definitions.)

| Variable | Definition |
|---|---|
| Degree Centrality | Number of direct friends/contacts an actor has |
| Betweenness Centrality | Probability that an actor is on the shortest path between any two other actors |
| In-group Degree | Number of direct friends from the same university |
| Out-group Degree | Number of direct friends from outside the same university |
| In-group Betweenness | Probability that an actor is on the shortest path between any two other actors from the same university |
| Out-group Betweenness | Probability that an actor is on the shortest path between any two other actors from different universities |
| Public/Private | 0/1 variable. Private =1. |
| Lawyer: degree centrality | Number of direct lawyer friends/contacts an actor has |
| Banker: degree centrality | Number of direct banker friends/contacts an actor has |
| Venture Capitalist: degree centrality | Number of direct VC friends/contacts an actor has |
| Entrepreneur Success 2009 | See Table 3 |
| Entrepreneur Success 2010 | See Table 3 |

*Table 2. Variables and Definitions*

We coded advisor and funding resources in each network, specifically identifying lawyers, bankers and venture capitalists and calculated the number of resources available to each entrepreneur.

**Entrepreneur success**

Finally, we classified the accomplishments of the entrepreneurs into five categories, based on differing levels of success (see Table 3 for the list of criteria). We evaluated startup success of each of the 960 founders twice, first in early 2009, and then 18 months later at the end of 2010, using the same framework. One researcher classified the success of each founder in 2009 and another researcher classified the success of the founders 18 months later. We looked up company Web sites on Google, also factoring in the number and importance of Google hits, and examined annual reports on the Hoppenstedt business database (www.hoppenstedt.de), if they were available.

| Success Level | Description |
|---|---|
| 1 | Company bankrupt / web site not existing / side business < 1 year |
| 2 | Company in business < 5 years / side business |
| 3 | Small size (2-5) > 5 years / main income / successful |
| 4 | Medium size (5-20) / family business / stable / very successful |
| 5 | Large size (>20) / successful projects / external funding / rewards |

*Table 3. Success categories for individual entrepreneurs*

Table 4 shows the number of entrepreneurs from the 12 universities in each of the original five success categories.

| University | Success Level 1 | Success Level 2 | Success Level 3 | Success Level 4 | Success Level 5 | Sum |
|---|---|---|---|---|---|---|
| EBS (private) | 0 | 3 | 44 | 29 | 4 | 80 |
| FU Berlin | 0 | 13 | 37 | 28 | 2 | 80 |
| HU Berlin | 0 | 22 | 42 | 14 | 2 | 80 |
| LMU Munich | 0 | 7 | 33 | 33 | 7 | 80 |
| RWTH Aachen | 2 | 18 | 41 | 19 | 0 | 80 |
| TU Munich | 0 | 19 | 48 | 10 | 3 | 80 |
| U Cologne | 1 | 15 | 42 | 20 | 2 | 80 |
| U Hamburg | 0 | 7 | 40 | 29 | 4 | 80 |
| U Hannover | 0 | 14 | 51 | 14 | 1 | 80 |
| U Karlsruhe | 0 | 10 | 37 | 29 | 4 | 80 |
| U Mannheim | 3 | 18 | 45 | 13 | 1 | 80 |
| WHU (private) | 2 | 7 | 34 | 32 | 5 | 80 |

*Table 4. Number of entrepreneurs from the 12 universities in each of the five success levels in 2009*

## 4. Results

For each university we calculated its alumni network (a proxy for an in-group), and the full network statistics. We then compared the network of the 960 entrepreneurs with the network of 960 randomly chosen non-entrepreneurs.

We calculated network variables twice, once for the full graph with 981,385 nodes, and once for the core graph (99,501 nodes, with degree centrality larger than or equal to 43, and 1,448,388 edges). The core graph included 300 of the randomly chosen entrepreneurs, but only 120 of the randomly chosen non-founders, giving a first indication that entrepreneurs on average potentially have much higher degree centrality than non-entrepreneurs.

| | | Degree | In-group degree | Out-group degree | In-group bc | Out-group bc |
|---|---|---|---|---|---|---|
| Entrepreneurs (N=960) | **mean** | **49.89** | **5.55** | **35.28** | **0.00160** | **0.00012** |
| | stdv | 64.93 | 9.24 | 45.68 | 0.00621 | 0.00000 |
| Non-entrepreneurs (N=960) | **mean** | **24.49** | **4.15** | **16.85** | **0.00088** | **0.00005** |
| | stdv | 24.28 | 7.36 | 18.22 | 0.00404 | 0.00000 |

*Table 5. T-test comparisons of networking variables between entrepreneurs and non-entrepreneurs. All differences are significant at the p < .05 level.*

We did a series of t-tests between entrepreneurs and non-entrepreneurs (see Table 5) to test Hypothesis 1. Analyses show there is a significant difference between entrepreneurs and non-entrepreneurs, with the entrepreneurs on average having twice as many links and being twice as embedded as the non-entrepreneurs. Comparing in-group (with their alumni) and out-group (with non-alumni) degree and betweenness, we find that although entrepreneurs still have higher in-group centrality, the major part of their network is with their out-group, this is in marked difference to the non-entrepreneurs. Entrepreneurs have twice the in-group and out-group betweenness centrality of non-entrepreneurs, meaning that they control information flow with alumni and external acquaintances, proving Hypothesis 1 – entrepreneurs have a much higher number of online friends than non-entrepreneurs and are more central in the overall network. Table 6 shows the Pearson correlation coefficients between our variables, only including the 809 actors who show up again 18 months later in our entrepreneurial success analysis. In addition, a partial reduction of the sample size was also due to a data quality issue that forced us to exclude TU Munich.

| | Variable | 1 | 2 | 3 | 4 | 5 | 6 | 7 | 8 | 9 | 10 | 11 | 12 |
|---|---|---|---|---|---|---|---|---|---|---|---|---|---|
| 1 | Success 2010 | 1.000 | | | | | | | | | | | |
| 2 | Success 2009 | .453** | 1.000 | | | | | | | | | | |
| 3 | Gender | -.030 | -.069 | 1.000 | | | | | | | | | |
| 4 | Degree | .051 | .138** | -.110** | 1.000 | | | | | | | | |
| 5 | Indegree | .054 | .141** | -.113** | .986** | 1.000 | | | | | | | |
| 6 | Outdegree | -.016 | -.011 | .007 | .149** | -.017 | 1.000 | | | | | | |
| 7 | Lawyers Outdegree | .085* | -.004 | .053 | .010 | -.030 | .243** | 1.000 | | | | | |
| 8 | Bankers Outdegree | .021 | -.011 | .008 | .032 | -.005 | .220** | .015 | 1.000 | | | | |
| 9 | Venture Cap. Outdegree | .030 | .072* | -.020 | .134** | .075* | .362** | .057 | .030 | 1.000 | | | |
| 10 | Lawyers Indegree | .071* | .058 | -.036 | .315** | .316** | .013 | .239** | .003 | .004 | 1.000 | | |
| 11 | Bankers Indegree | .046 | .046 | -.054 | .569** | .577** | -.008 | -.016 | .137** | .074* | .158** | 1.000 | |
| 12 | Venture Cap. Indegree | .074* | .099** | -.064 | .649** | .667** | -.064 | -.021 | -.014 | .137** | .160** | .544** | 1.000 |

* $p < .05$, ** $p < .01$.

*Table 6. Person correlation coefficient for variables at Level 1 (N=809).*

As Table 6 illustrates, success in 2009 – at the same time when the XING links were collected – is correlated with degree, both with the in-group and the out-group. This suggests that having many friends is an indicator of current success of an entrepreneur, however, we do not observe such an effect for success in the following year. The same is true for venture capitalists: the more venture capitalist friends an entrepreneur has, the more successful they are concurrently. We do not see such a concurrency effect for lawyers. Rather, having many lawyers as XING friends in 2009 seems to predict future success of an entrepreneur in 2010. Perhaps lawyers are more risk-averse and are better in identifying success of an entrepreneur in the subsequent year. Venture capitalists, with their risk-taking reputation, are positively correlated with success the same year. The venture capitalist in-group also seems to be able to predict success, perhaps because they have more information from their fellow alumni entrepreneurs. The number of bankers as XING friends, on the other hand, does not correlate with success. The gender of entrepreneurs has no influence on their success in our sample. It therefore seems that having many lawyers as online friends might predict next year entrepreneurial success.

| Variable | Model 1 | Model 2 | Model 3 | Model 4 | Model 5 | Model 6 | Model 7 |
|---|---|---|---|---|---|---|---|
| Constant | 26.141** | 26.153** | 26.265** | 26.141** | 26.143** | 26.135** | 26.135** |
| **Level 1 - Students** | | | | | | | |
| Success 2009 | | 6.681** | | | | | 6.640** |
| Gender | | | -.944 | | | | |
| Indegree | | | | .009 | | | |
| Outdegree | | | | -.017 | | | |
| Lawyers Outdegree | | | | | 1.118* | | 1.265** |
| Bankers Outdegree | | | | | .753 | | |
| Venture Cap. Outdegree | | | | | .352 | | |
| Lawyers Indegree | | | | | .183 | | |
| Bankers Indegree | | | | | -.006 | | |
| Venture Cap. Indegree | | | | | .817 | | |
| **Level 2 - Universities** | | | | | | | |
| University Size | | | | | | $-6.75 \times 10^{-5*}$ | $-6.79 \times 10^{-5**}$ |
| **Random-effects parameters** | | | | | | | |
| Variance Level 2 | .481 | .615 | .428 | .462 | .153 | $2.15 \times 10^{-12}$ | $5.80 \times 10^{-20}$ |
| Variance Level 1 | 127.538 | 101.165 | 127.480 | 127.156 | 125.917 | 127.043 | 99.826 |
| ICC | 0.38% | | | | | | |
| Change in variance Lev. 1 | | 27.85% | -11.05% | -3.93% | -68.24% | -100.00% | -100.00% |
| Change in variance Lev. 2 | | -20.68% | -.05% | -.30% | -1.27% | -.39% | -21.73% |
| N | 809 | 809 | 809 | 809 | 809 | 809 | 809 |
| Groups | 11 | 11 | 11 | 11 | 11 | 11 | 11 |

*$p < .05$, ** $p < .01$. ICC = Intraclass correlation coefficient.

*Table 7. Predicting entrepreneurs' success in 2010.*

Table 7 shows the multilevel regression models, using full maximum likelihood estimation, with the predictors centered on their sample mean. The first six models assess the contribution of each subset of predictors; all significant predictors are included in model 7. Not surprisingly, the strongest predictor of success in 2010 is success in 2009. We find that the number of friends in 2009, i.e. the degree of entrepreneurs, as well as their gender, does not have a significant impact on their entrepreneurial success (nor does betweenness, not shown in the regression). Also having many connections to bankers or venture capitalists does not help. What is important to achieve success in 2010 is having been successful the previous year. Moreover, coming from a small university and having more connections with lawyers will be of very small but significant help.

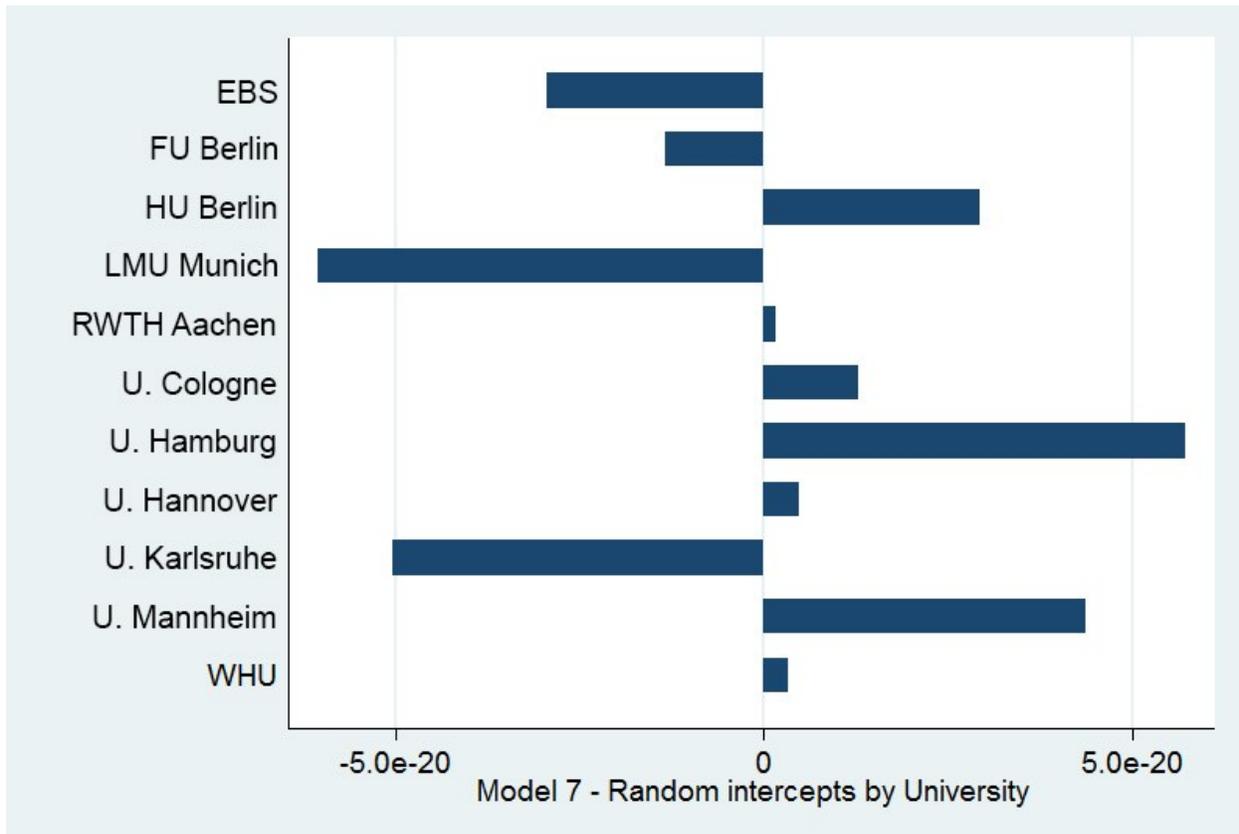

*Figure 3.  Average success in 2010 for alumni from each university*

Figure 3 reveals that, for a given level of success in 2009, lawyer outdegree and university size, for students from U. Hamburg compared with those from LMU Munich, the success score in 2010 is, on average, about $10^{-19}$ points higher. In other words, the difference attributable to the university is extremely small. The intraclass correlation coefficient (ICC) of 0.38% is close to 0, which means that a very small proportion of the variance in entrepreneurial success is attributable to universities (Figure 3): most of the variance in success is due to individual characteristics of the entrepreneurs, and it does not matter which university somebody graduated from in Germany. Note that a low ICC does not mean that the multilevel structure of the data can be ignored (Nezlek, 2008). Rather it means that it matters very little for the entrepreneurs in Germany if they study at a public or a private university, the only difference is if the university is large or small (small is slightly better). The key finding of the correlations in Table 6 and the multilevel regression is that indeed the number of friends in XING has no influence on business success in the following year.

| | virtual ties | real world ties |
|---|---|---|
| network size | size does not matter | size matters (Raz & Gloor 2007) |
| embeddedness | embeddedness does not matter | embeddedness matters (Uzzi 1996) |
| location | location matters somewhat | location matters (Owen-Smith&Powell 2004) |
| diversity | Diversity matters somewhat | Diversity matters (Gilsing & Noteboom 2005) |

Figure 4. Research Findings

Figure 4 resumes our research findings, contrasting them with previous research on real-world ties in the brick-and-mortar world.

## 5. Discussion and Conclusion

In this project, we tackle the research questions whether there is difference between real-world and virtual ties, and under what circumstances these virtual ties might create business benefits for entrepreneurs. Extending previous work in correlating social network structure and embeddedness in the real world with success (e.g., Cummings and Cross, 2003; Raz and Gloor, 2007; Uzzi, 1996; 1997) to online social networking, we have shown that structural properties of online social networks of entrepreneurs is immaterial to the success of those entrepreneurs. It seems that online ties are only as good as the real-world relationships that underlie them. In general, it is difficult to extract the quality of a real-world tie from the social signal on Facebook or LinkedIn. The context of the link gives some cues about tie strength, such as a link between two alumni of the same university. If this university is small, chances are the tie is somewhat stronger than if the two alumni are from a large university (see Table 2). Small universities have a different virtual networking structure, i.e., the network is much smaller, and the density and closure is higher (Wasserman and Faust, 1994), suggesting two alumni are more likely to know each other in the real world. Our findings are consistent with Chen (2013) who stresses the importance of making a distinction between weak and strong ties in online communication. Similarly, Haythornthwaite (2002) found that the strength of social ties differs offline and online and that communication on social media has a positive impact on both strong and weak ties.

The main conclusion for entrepreneurs is: accumulating many online social networking ties does not help for entrepreneurial success. It does not help to befriend everyone who might accept an invitation.

There seem to be significant differences in weak tie links in the real world compared to the weak ties in the public online world (in networks like Facebook or XING and LinkedIn). Weak ties in the face-to-face world are often conduits for information. We saw no evidence of value being created, however, in weak tie links in the online world. Our research suggests that public networks with their many weak ties are just a place to aggregate contact information.

There are ties, however, that alumni of universities can cultivate to increase the chance of success. The number of ties to venture capitalists was not significant to the success for either group of founders; the same was true for bankers. The fact that ties to venture capitalists made no difference to success may reflect different new venture funding mechanisms in North America and Germany. However, access to lawyers – what we call an advisory resource – is significant to the success of founders. This suggests that founders should cultivate people for their network who have specific skills and have access to information.

As Porter et al. (2005) demonstrated, most university-educated founders retain some form of affiliation with their universities after successfully starting their business. One recommendation for German universities therefore might be to re-create the stratified small-group elitist environment of US private universities like Harvard or MIT. This is something that cannot be done overnight. The value of the alumni network of a US elite university depends on members who have successfully grown their startup companies, or have risen to senior management levels and board rooms of large companies. Public German universities could try to re-create similar "entrepreneurship clubs" with a high barrier to entry, and access to top-level power brokers among their alumni. Our results indicate that entrepreneurs, especially from large public universities, should not waste their time building a large outside online social network at all cost, as investing into acquiring many links seems to take away focus from building up the business.

Finally, in this type of analysis, it is hard to infer causality. Are entrepreneurs successful because they have many alumni friends, or do they have many alumni friends because they are successful?  We address causality in two ways. First, we tried to disentangle the success measure from the friendship network by measuring success 18 months after we measured the online network, giving at least an indication if having the right type of links now might lead to success in the future. Second, we assume that business founders and entrepreneurs are more likely to want an online presence (at least in the business setting which is why we focused on XING, a business networking site). This implies that we can expect to obtain more links from and to entrepreneurs than non-entrepreneurs. This has indeed been shown. However, there is little difference between successful and unsuccessful entrepreneurs in the number of friends. We suspect that collecting an inordinate number of online links is an attempt to create a positive signal of popularity, similar to Justin Bieber and Lady Gaga having the most Twitter followers thanks to their real-world popularity. While for Justin Bieber and Lady Gaga the links are a sign of real-world popularity, we suspect that for most people causality is the other way around: collecting links in an attempt to increase the strength of their signal for the real world.

**Limitations and Future Research**

This study has a number of possible limitations. The main problem is potential confounding effects between exogenous and endogenous variables, as success could also have an impact on social capital, assessed in this study as the number of friends. As has been discussed above, this means that a reverse causality bias is likely. Second, there might be personality characteristics of the entrepreneurs like the ability to communicate that could have an impact on both social capital and success, thus introducing an omitted variable bias. As we do not have direct personality characteristics assessments of entrepreneurs, we are unfortunately unable to directly address this question.

Another possible limitation of this study is that there could be a selection bias. It is possible that our Internet-based sample might not be relevant for the entire population of German founders. Since not everybody participates in all online social networks, we can only make conclusions about the XING alumni population. We believe, however, that our sample is especially representative of the entire younger founder population of Germany. These entrepreneurs, the Web savvy age group of twenty to forty year olds, have a high likelihood of using tools like XING to stay in touch. We think that it is also likely that older entrepreneurs are using social networks like XING to keep track of their contact. We might miss some blue-collar startup founders such as butchers, hairdressers, or carpenters, who might not be on XING, but these are not part of our study, as the emphasis of our project is on high-growth technology-oriented startup companies.

Another limitation is that these results may not generalize well. We did compare network demographics to other professional networks like LinkedIn and the XING sample we used was similar. However, the German university system may not be comparable to university systems in other countries. The German university landscape may be much simpler than the university landscape in the U.S., as the U.S. has an enormous variety of universities. Using the German university system as our sample does allow us to make strong in-group, out-group predictions.

A third limitation is the paucity of additional demographic and other information that could deepen our understanding of in-group/out-group relationships to business success. This is an outcome of the data-gathering design, but it could be addressed in future work.

The main implications for further research are based on the insight gained in this study that online social networking ties "per se" are weak indicators of real-world networks. While it has been found that all types of real-world networks are important for entrepreneurial success, our findings indicate that online social networking ties are weak correlates of entrepreneurial success. We therefore call on other researchers to develop more reliable indicators of real-world ties from online social network cues. Such cues might be shared attributes, such as being an alumnus of the same university, having worked at the same company, or coming from the same geographical region. We also speculate that there might be more subtle indicators of success, such as non-obvious dynamic communication patterns obtained through analyzing e-mail archives, or Twitter records. For early results see for instance (Gloor, 2016; Gloor et al., 2014).

Another area for further research will be to explore what type of business the founders have been starting: it might be that for some types of businesses social networking ties are more important than for others. Lastly one could study the motivation for creating links. We rely on the assumption that entrepreneurs are making calculations about the links they accept and request but, in the absence of research like in-depth interviews, those are only assumptions.

**Conclusion**

We do not find that having many online social networking friends adds to the future business success of entrepreneurs. Taking the alumni networks of ten German universities as a proxy for location-based sub-clusters, we find that acquiring many friends through a business network like LinkedIn or XING does not seem to help entrepreneurs. It looks as if there are better ways for entrepreneurs to spend their time, rather than amassing many online friends. Yes, entrepreneurs appear more popular if they have more online friends, but unfortunately this does not help them to become more popular and successful in the real world.